\documentclass[12pt,a4paper]{article}
\usepackage[english]{babel}
\usepackage[latin1]{inputenc}
\usepackage[T1]{fontenc}
\usepackage{amsmath}
\usepackage{amsfonts}
\usepackage{amssymb}
\usepackage[dvips]{graphicx,color}
\usepackage{slashed}
\usepackage{physics}
\begin{document}
\title{Chiral anomaly, induced current, and vacuum polarization tensor for a Dirac
field in the presence of a defect} 
\author{C.~D.~Fosco and
A.~Silva\\
{\normalsize\it Centro At\'omico Bariloche and Instituto Balseiro}\\
{\normalsize\it Comisi\'on Nacional de Energ\'{\i}a At\'omica}\\
{\normalsize\it R8402AGP S.\ C.\ de Bariloche, Argentina.} }
\maketitle
\begin{abstract} 
	We evaluate the vacuum polarization tensor (VPT) for a massless
	Dirac field in $1+1$ and $3+1$ dimensions, in the presence of a
	particular kind of defect, which in a special limit imposes bag
	boundary conditions. 

	We also show that the chiral anomaly in the presence of such a
	defect is the same as when no defects are present, both in $1+1$
	and $3+1$ dimensions. This implies that the induced vacuum current in
	$1+1$ dimensions due to the lowest order VPT is exact.
\end{abstract}
\maketitle
Chiral and other anomalies have been objects of intense research, since the
pioneering derivation of the anomalous divergence of the axial
current~\cite{adleranomaly}.  Their relevance to diverse theoretical and
phenomenological aspects of Quantum Field Theory (QFT) can hardly be
emphasized.  This is one of the reasons why the chiral anomaly has been
derived in the context of many diverse models, and by following different
approaches~\cite{fujikawaoriginal,pointsplitting,jackivtriangle} (for a
comprehensive  review of chiral and other types of anomalies see, for
example~\cite{resumebilial}). 

In physical terms, anomalies have their origin in the existence of UV
divergences: short-distance fluctuations that require the use of a
regulator.  This, either implicitly or explicitly, implies the introduction of a {\em mass\/} scale $\Lambda$, which results in the breaking of a symmetry
that depends on the absence of any such scale. This breaking manifests
itself in the violation of a classical conservation law, a violation which survives
 the removal of the regulator, i.e., \mbox{$\Lambda \to \infty$}.

On the other hand, introducing nontrivial boundary conditions in QFT models
is a rather explicit breaking of (at least) translation symmetry. This
leads to  many interesting effects, a noteworthy example of which being the
Casimir effect~\cite{Casimir:1948dh,Bordag:2001qi}, as well as many other
related phenomena~\cite{Milton:2001yy}. In the case of fermionic fields, non-trivial boundary conditions, and the resulting Casimir effect, has been studied extensively within the bag model of QCD~\cite{Chodos:1974pn}. Besides, interesting applications of the fermionic Casimir effect to carbon nanotube models have been presented~\cite{Bellucci:2009hh,Elizalde:2011cy}. 
It is the aim of this letter to study the interplay between the two
phenomena, anomalies and non trivial boundary conditions, in a concrete system: fermions in the presence of a potential, which may be used to impose bag boundary conditions. As shown in~\cite{Saghian:2012zy}, in order to impose that kind of condition, the potential has to be a singular, space-dependent mass term.

We analyze the results of that interplay on two objects: the chiral anomaly
and the vacuum polarization tensor (VPT), a correlator between current
fluctuations, which one should expect to exhibit a strong dependence with
the distance to the boundary. 

The structure of this work is as follows: in Sect.~\ref{sec:themodel} we
introduce the model; then, in Sect.~\ref{sec:chiral} we
evaluate the chiral anomaly for that model, and in Sect.~\ref{sec:vpt} the
induced vacuum current and the VPT.  
Finally, in Sect.~\ref{sec:conc} we present our conclusions.

\section{The model}\label{sec:themodel}
We consider a quantum fermionic field ($\psi, \, \bar\psi$) in $D=d+1$
dimensions \mbox{($d=1,\,3$)}, endowed with an Euclidean action
$S(\bar\psi,\psi;A)$, with $A$ an external Abelian gauge
field. We impose non-trivial boundary conditions on the fermionic field at
$x_d=0$, by coupling it to a singular scalar potential:
\begin{equation}\label{eq:defs}
{\mathcal S}({\bar\psi},\psi;A)\;=\;
	\int d^{d+1}x \,{\bar\psi}(x) \, \big[ \not \!\! D  + g
	\delta(x_d) 
\big] \, \psi(x) \;,
\end{equation}
 with $\not \!\! D = \not \! \partial + i e \not\!\! A(x)$,  and $g$ a dimensionless constant.  
It is worth noting that in~\cite{Fuentes:1995ym} a $1+1$ dimensional model has been studied, where a singular potential is part of the gauge potential; namely, $A_{\mu}(x) = B_{\mu}(x) + s_{\mu}(x_{0})\delta(x_{1})$, with $B_{\mu}$ a bulk contribution of the gauge potential, and $s_{\mu}$ a singular term. 

Recalling the approach of~\cite{EfeCasFer,Ttira:2010rh}, one sees that
$g=2$ imposes {\em bag\/} boundary conditions~\cite{EfeCasFer}, namely, the
normal component of the vector current $J_\mu$ due to the Dirac field,
vanishes on the interface $x_d=0$. We assume that the fermions are confined
to the region which, in our choice of coordinates, corresponds to $x_d >0$.
For $g=2$, it becomes disconnected (independent) from its complement. For
$g \neq 2$, however, that is no longer true, as part of the current may
cross the interface between them.

In our conventions, both $\hbar$ and the speed of light are equal to $1$,
spacetime coordinates are denoted by
$x_\mu$, $\mu\,=\,
0,\,1,\, \ldots, d$, and the metric tensor is \mbox{$g_{\mu\nu} \equiv {\rm
diag}(1,1, \ldots, 1)$}. Regarding Dirac's $\gamma$-matrices, for $d=1$ they are chosen in the representation:
\begin{equation}\label{eq:gamma_matrices}
\gamma^0 \,\equiv\, \sigma_1 \,=\, 
\left(
\begin{array}{cc}
	0 & 1 \\
	1 & 0
\end{array}
\right)
\;,\;\;
\gamma^1 \,\equiv\, \, \sigma_2 \,=\, 
\left(
\begin{array}{cc}
0 & -i \\
i & 0
\end{array}
\right) \;,
\end{equation}
and 
\begin{equation}
\gamma^5 \,\equiv\, \gamma_5 \,\equiv\, -i  \gamma^0 \gamma^1 \,=\, 
\sigma_3 \,=\,\left(
\begin{array}{cc}
1 & 0 \\
0 & -1
\end{array}
\right) \,,  
\end{equation}
with $\sigma_i$ ($i=1,\,2,\,3$) representing the standard Pauli's matrices. On the other hand, for $d=3$:
\begin{equation}
	\gamma_\mu \;=\; \left( \begin{array}{cc} 0 & \sigma_\mu^\dagger \\ 
	\sigma_\mu & 0 \end{array}\right) \;\;,\;\;\;
	\gamma_5 \;\equiv\; \gamma_0 \gamma_1 \gamma_2 \gamma_3 \,=\, \left(
	\begin{array}{cc} {\mathbb I}_{2\times 2} & 0 \\ 
	0 & - {\mathbb I}_{2\times 2} \end{array}\right) \;,
\end{equation}
where $\sigma_0 \equiv i {\mathbb I}_{2\times 2}$.

The boundary conditions that this system imposes on
the fields are as follows: the singular potential in the Dirac equation implies a
discontinuity in $\psi$ which, following~\cite{Fosco:2007ry}, may be
replaced by the average
of the two lateral limits:
\begin{equation}\label{eq:disco1}
 \gamma_d \big( \psi(x_\shortparallel, \epsilon) -
\psi(x_\shortparallel,-\epsilon) \big) \,+\,  \frac{g}{2} 
\big( \psi(x_\shortparallel,\epsilon) +
\psi(x_\shortparallel,-\epsilon) \big)  \;=\;0 \;,
\end{equation}
with $x_\shortparallel \equiv (x_0,x_1,\ldots,x_{d-1})$.

Setting $g=2$, and introducing the (orthogonal) projectors: ${\mathcal
P}^{\pm} \equiv \frac{1 \pm \gamma_d}{2}$, 
\begin{equation}
{\mathcal P}^+ \psi(x_\shortparallel, \epsilon)  \;=\; - \, 
{\mathcal P}^- \psi(x_\shortparallel, -\epsilon)  \;.
\end{equation}
Thus, the orthogonality of these projectors leads to:
\begin{equation}
{\mathcal P}^+  \psi(x_\shortparallel, \epsilon)  \;=\;  0 \;\;,\;\;\;\;
{\mathcal P}^-  \psi(x_\shortparallel, -\epsilon)  \;=\;0 \;.
\end{equation}
Each one of these conditions implies the vanishing of $\bar{\psi}(x)\gamma_d \psi(x)$, the normal component of the vacuum current (see next Section below) for $g=2$, approaching the border, or `wall' either from $x_d >0$ or from $x_d < 0$. 
\section{Chiral anomaly in the presence of a defect}\label{sec:chiral}
Let  $j_\mu^5(x) \equiv \langle J_\mu^5(x) \rangle$ the vacuum
expectation value of the axial current 
$J_\mu^5(x) \equiv i e  \bar{\psi}(x)\gamma_\mu \gamma_5 \psi(x)$, 
where the average symbol $\langle \ldots \rangle$ is defined as follows:
\begin{equation}
\langle \, \ldots \,  \rangle \;\equiv \; \frac{\int {\mathcal D}\bar{\psi} {\mathcal
D}\psi \ldots e^{-S(\bar{\psi},\psi;A)}}{\int {\mathcal D}\bar{\psi} {\mathcal
D}\psi \, e^{-S(\bar{\psi},\psi;A)}} \;,
\end{equation}
with $S$ as in (\ref{eq:defs}). 
A {\em naive\/} evaluation of the  divergence of $j_\mu^5(x)$, using
the equations of motion satisfied by the Dirac field, yields the wrong
(classical) result:
$\partial_\mu j_\mu^5(x) = 2 i e  g  \delta(x_d) \langle \bar{\psi}(x)
\gamma_5 \psi(x) \rangle$. This is wrong because of the
ill-defined nature of the fermion bilinear. A possible way to tackle this
in a gauge invariant manner is to use a (single), bosonic, Pauli-Villars regulator
field $\phi$, with a mass $\Lambda$, having a Dirac action with the same
couplings as the Dirac field. This produces for the
divergence~\footnote{The average symbol for the regulator field is defined
in an entirely analogous fashion as for the original field, except for its
mass and opposite statistics.}:
\begin{equation}
	\partial_\mu j_\mu^5(x) = \lim_{\Lambda \to \infty}\left\{  
	2 i  e g  \delta(x_d) \big[ \langle \bar{\psi}(x)\gamma_5 \psi(x) \rangle 
	+ \langle \bar{\phi}(x)\gamma_5 \phi(x) \rangle \big] 
	\,+\, \Lambda \langle \bar{\phi}(x)\gamma_5 \phi(x) \rangle
	\right\} \;.
\end{equation}
Therefore
\begin{equation}
\partial_\mu j_\mu^5(x) = {\mathcal A}(x) \,+\,	2 i  e g
\delta(x_d) \, \langle \bar{\psi}(x)\gamma_5 \psi(x)\rangle 
\end{equation}
where
\begin{equation}\label{eq:ganom}
{\mathcal A}(x) \;=\;
 2 i e \, \lim_{\Lambda \to \infty} \,
	(\Lambda + g \delta(x_d)) {\rm tr} \left[\gamma_5
	\langle x| \big( \not\!\! D  + g \delta(x_d) + \Lambda \big)^{-1} |x\rangle  
	\right] \,
\end{equation}
where Dirac's bra-ket notation has been used to denote matrix elements of
the inverse of the Dirac operator (which includes the singular potential
and the gauge field). 
At this point we note that, for $g=0$ (i.e., no defect), one can show that:
\begin{equation}
{\mathcal A}|_{g=0}(x) \;\equiv\; {\mathcal A}_0(x) \;=\; 2 i e
\lim_{\Lambda \to \infty} \, {\rm tr} 
	\Big[\langle x| f(-\frac{\not \!\! D^2}{\Lambda^2}) |x\rangle \Big]
\end{equation}
where $f(x) \equiv \frac{1}{1+x}$. Since $f$ is a function which satisfies
$f(0)=1$, and $\lim_{x \to \infty} f^{(k)}(x) = 0$ for all $k$, this
reproduces the proper result for the anomaly, namely, 
\begin{equation}
 {\mathcal A}_0(x)  = \frac{2 i e^{1+\frac{D}{2}}}{
 (4\pi)^{\frac{D}{2}}(\frac{D}{2})!}\epsilon_{\mu_{1}\nu_{1}...\mu_{\frac{D}{2}}\nu_{\frac{D}{2}}}F_{\mu_{1}\nu_{1}}(x)...F_{\mu_{\frac{D}{2}}\nu_{\frac{D}{2}}}(x),
\end{equation}
where $\epsilon_{\mu_{1}\nu_{1}...\mu_{\frac{D}{2}}\nu_{\frac{D}{2}}}$ is
the Levi-Civita Symbol in $D=d+1$ dimensions. 

The anomaly ${\mathcal A}$ can also be obtained in terms of the
anomalous Jacobian $J_{\phi}$  due to the (infinitesimal version of the)
transformation
\mbox{$\psi(x) \rightarrow e^{i e  \phi(x)\gamma_{5}} \psi(x)$},
${\bar\psi}(x) \rightarrow {\bar\psi}(x) e^{i e  \phi(x)\gamma_5}$,
as follows:
\begin{equation}
{\mathcal A}(x) \;=\; 2 i e \frac{\delta \log[J_\phi]}{\delta\phi(x)}
\;,
\end{equation}
where
\begin{equation}
\log[J_{\phi}] \equiv \lim_{t\to0^{+}}\sum_{k=0}^{\infty} t^{\frac{k-D}{2}} a_{k}(\phi\gamma^{5},\slashed{D}^{2} ),
 \label{eq:3.5}
\end{equation}
where $a_{k}$ are functions of matrix-valued arguments. Let us consider now the form of the anomaly for two cases: bag boundary conditions and the general ($g$ not necessarily equal to $2$):
\subsection{Bag boundary conditions}
For the case of bag boundary conditions (in our set-up: $g=2$), and  more general
geometries, the calculation of the coefficients in (\ref{eq:3.5}) has been presented
in~\cite{anomalybagconditions}.
For a planar wall, the first $5$ coefficients reduce to: 
\begin{align}
     a_{0}(\phi\gamma^{5},\slashed{D}^{2} ) &= (4\pi)^{\frac{-D}{2}}
	(\int_{x_d > 0}d^{D}x \, {\rm tr}(\phi(x)\gamma_{5})),
    \stepcounter{equation}\tag{\theequation}\label{eq:3.6}
    \\ 
    a_{1}(\phi\gamma_5,\slashed{D}^{2} ) &=
	\frac{1}{4}(4\pi)^{\frac{-d}{2}}( \int d^d x_\shortparallel \,  {\rm
	tr}(\chi \phi(x_\shortparallel,0)\gamma_{5})),
    \stepcounter{equation}\tag{\theequation}\label{eq:3.7}
    \\ 
    \nonumber a_{2}(\phi\gamma_{5},\slashed{D}^{2} ) &=
	\frac{1}{6}(4\pi)^{\frac{-D}{2}}( \int_{x_d > 0}d^{D}x \,
	{\rm tr}(6\phi(x)\gamma_{5} [\gamma_{\nu},\gamma_{\mu}]
	\frac{F_{\mu\nu}(x)}{4}) 
	\\\nonumber&+ \int d^dx_\shortparallel {\rm tr}( 3\chi
	\partial_{d}\phi(x_\shortparallel,0)\gamma_{5} )   ),
    \stepcounter{equation}\tag{\theequation}\label{eq:3.8}
    \\ 
    \nonumber a_{3}(\phi\gamma_{5},\slashed{D}^{2} ) &= 
\frac{1}{384}(4\pi)^{\frac{-(D-1)}{2}}(\int dx_\shortparallel
	{\rm tr}(\phi(x_\shortparallel,0)\gamma^{5}(96\chi
	[\gamma_{\nu},\gamma_{\mu}] \frac{F_{\mu\nu}(x_\shortparallel,0)}{4} ) 
    \\\nonumber&+ 24\chi \partial_{d}\partial_{d}\phi(x_\shortparallel,0)\gamma^{5} )),
    \stepcounter{equation}\tag{\theequation}\label{eq:3.9}
    \\
     \nonumber a_{4}(\phi\gamma^{5},\slashed{D}^{2} ) &=
	\frac{1}{360}(4\pi)^{\frac{-D}{2}}(\int_{x_d > 0}d^{D}x \,
	{\rm tr}(\phi(x)\gamma^{5}( 60
	\partial_{i}\partial_{i}[\gamma_{\nu},\gamma_{\mu}]
	\frac{F_{\mu\nu}(x)}{4}
     \\ \nonumber & + 180 E^{2}(X) -30F_{\mu\nu}^{2}))
    \\\nonumber
	&+  \int dx_\shortparallel \, {\rm tr}(\phi(x)\gamma_{5}( (240\sqcap_{+} -120\sqcap_{-})\partial_{d}[\gamma_{\nu},\gamma_{\mu}] \frac{F_{\mu\nu}(x)}{4} )
    \\ \nonumber
    &+ \partial_{d}\phi(x)\gamma_{5}( 180\chi [\gamma_{\nu},\gamma_{\mu}]
	\frac{F_{\mu\nu}(x)}{4} ) +
	30\partial_{i}\partial_{i}\partial_{d}\phi(x)\gamma_{5}\chi ) ),
    \stepcounter{equation}\tag{\theequation}\label{eq:3.10}
\end{align}
where $\sqcap_{+} \equiv \frac{1}{2}(1+i\gamma^{5}\gamma_{d})$, $\sqcap_{-}
\equiv \frac{1}{2}(1-i\gamma^{5}\gamma_{d})$, $\chi \equiv \sqcap_{+} -
\sqcap_{-}$ and $\partial_{i}$ denotes derivation with respect to all
coordinates except $x_{d}$.
Higher-order coefficients are
multiplied by a positive power of $t$ in the expansion \ref{eq:3.5}, so
they do not contribute. 

In $D=2$, just the coefficients \ref{eq:3.6},
\ref{eq:3.7} and \ref{eq:3.8} are relevant. Given that the trace of an odd
number of $\gamma$ matrices vanishes \cite{greinerlibro}, and taking into
account the Dirac algebra, one sees that the contribution from $a_1$
vanishes. 

The coefficient $a_{2}(\phi\gamma^{5},\slashed{D}^{2})$ contains both
boundary ($x_1=0$) and bulk ($x_1 > 0$) terms. Only the latter is
non-vanishing, and it produces the anomaly in $2$ dimensions 
\begin{align}
{\mathcal A}(x)  &= \frac{i e^2}{2\pi} \epsilon_{\mu\nu}
F_{\mu\nu}(x)  
\stepcounter{equation}\tag{\theequation}\label{eq:3.15}
\end{align}
where $\epsilon_{\mu\nu}$ is the $2$-dimensional Levi-Civita tensor. Thus,
there is no quantum correction to the anomaly in $2$
dimensions when confining fields to the half-line. 

In $4$ dimensions,  $a_{0}$ and $a_{1}$ are zero, as well as 
the boundary term in $a_{2}$.  The bulk term is proportional to
$tr(\gamma^{5} \gamma_{\mu}\gamma_{\nu})$, which is a trace of $6$ $\gamma$
matrices. This trace is evaluated in more detail in \cite{trazagamma}, and
is equal to zero. Similarly to the precedents coefficients, $a_{3}$ has two
terms and they are proportional to $tr( \gamma^{5} \chi
\gamma_{\mu}\gamma_{\nu} )$ and $ tr( \gamma^{5} \chi \gamma^{5})$
respectively. Both of them are zero because they are traces of $11$ and
$13$ gamma matrices.

We are left with the contributions of the coefficient $a_{4}$. The boundary
contribution has $4$ terms. These are all proportional to the trace of an
odd number of $\gamma$ matrices or proportional to $tr(\gamma^{5}
\gamma_{\mu}\gamma_{\nu})$, so there is no quantum boundary contribution.
The bulk term yields the usual result
\begin{align}
{\mathcal A}(x)  &= \frac{i e^{3}}{16\pi^{2}}
\epsilon_{\mu\nu\alpha\beta} F_{\mu\nu}(x)F_{\alpha\beta}(x) 
\end{align}
where $\epsilon_{\mu\nu\alpha\beta}$ is the Levi-Civita tensor in $4$ dimensions.

Thus, so far we have proved that axial anomaly has no boundary contributions
when bag boundary conditions are imposed at $x_{d}=0$.
\subsection{General case}
In this case, we come back to the general expression (\ref{eq:ganom}) for
${\mathcal A}$, and note that the inverse operator appearing there may be
rendered as follows:
\begin{align}\label{eq:dexp}
\langle x| \big( \not\!\! D  + g \delta(x_d) + \Lambda \big)^{-1} |y\rangle 
	&=\, \langle x| \big( \not\!\! D  + \Lambda \big)^{-1} |y\rangle
	\nonumber\\
	 -  g \, \int d^dz_\shortparallel d^dz'_\shortparallel \, \langle x|
	\big( \not\!\! D  + \Lambda
	\big)^{-1}|z_\shortparallel, 0\rangle
	& M(z_\shortparallel,z'_\shortparallel)
	\langle z'_\shortparallel,0| \big( \not\!\! D  + \Lambda \big)^{-1} |y\rangle
\end{align}
with 
\begin{equation}
	M(z_\shortparallel,z'_\shortparallel) \;=\; \langle
	z_\shortparallel, 0 |\big[1 + g   (\not\!\! D  + \Lambda)^{-1}
	\big]^{-1}
	|z'_\shortparallel, 0  \rangle \;.
\end{equation}
Now, we see that the first term in (\ref{eq:dexp}) reproduces the previous,
bag model anomaly. The second term, whenever one considers points in the
bulk $x_d = \varepsilon > 0$ will produce a vanishing contribution when
$\Lambda \to \infty$. The reason is that that contribution is UV finite for
$\varepsilon > 0$, since one needs to consider that term for $y=x =
(x_\shortparallel, \varepsilon)$, and there are no coincident points in any 
operator. Therefore, the whole contribution is finite, and, since it has an extra
negative power of $\Lambda$, it vanishes in the limit.

\section{Induced vacuum current and vacuum polarization
tensor}\label{sec:vpt}
\subsection{Induced vacuum current}
Let  $j_\mu(x) \equiv \langle J_\mu(x) \rangle$ the vacuum
expectation value of the (vector) current, which is given by
\begin{equation}
j_\mu(x) \;=\; - \, e \, {\rm tr} \left[\gamma_\mu 
\langle x| \big( \not \!\partial + i e \not\!\! A(x) + g \delta(x_d) 
\big)^{-1} |y\rangle \ \right] \;.
\end{equation}

In order to express $j_\mu$ in terms of the VPT, we
expand that inverse in
powers of $A$: 
\begin{equation}
\langle x| \big( \not \!\partial + i e \not\!\! A + g \delta 
\big)^{-1} |y\rangle \,=\, {\mathcal S}_F(x,y) \,-\, \int d^{d+1}z \,
{\mathcal S}_F(x,z) \,i e \not\!\! A(z) \,{\mathcal S}_F (z,y)  \,+\,
	\ldots
\end{equation}
where ${\mathcal S}_F(x,y)$ is the exact propagator in the presence
of the defect, and no $A$.   
Therefore, to the lowest non-trivial order, we obtain:
\begin{equation}\label{eq:j1}
	j_\mu(x) \;=\; i\,e^2 \, \int d^{d+1}y \;  
	{\rm tr} \Big[\gamma_\mu \; {\mathcal S}_F(x,y) \gamma_\nu \;
	{\mathcal S}_F(y,x) \Big] \, A_\nu(y) \;.
\end{equation} 
Namely, to this order, the response to the external gauge field is linear, the
proportionality being given by the VPT, $\Pi_{\mu\nu}$, defined by:
\begin{equation}\label{eq:defpi}
\Pi_{\mu\nu}(x,y) \;=\; - \,e^2 \,  
	{\rm tr} \left[\gamma_\mu \; {\mathcal S}_F(x,y) \gamma_\nu \;
	{\mathcal S}_F(y,x) \right] \;,
\end{equation}
such that, 
\begin{equation}
	j_\mu(x) \;=\; - i \, \int d^{d+1}y \, \Pi_{\mu\nu}(x,y)  \, A_\nu(y) \;.
\end{equation}
\subsection{Vacuum polarization tensor in the presence of the
defect, in $1+1$ dimensions}\label{ssec:pi2}
Since the defect is static, ${\mathcal S}_F$, and therefore also
$\Pi_{\mu\nu}$, will depend on the time arguments only through their
difference.  Using a mixed Fourier representation whereby we just transform
the time coordinate,
\begin{equation}
	{\mathcal S}_F(x_0,x_1;y_0,y_1) \;=\; {\mathcal S}_F(x_0-y_0;x_1,y_1) 
\;=\;\int \frac{dp_0}{2\pi} e^{i p_0 (x_0-y_0)} \;
	\widetilde{\mathcal S}_F(p_0;x_1,y_1) \;, 
\end{equation}
we see that:
\begin{equation}\label{eq:pi}
	\widetilde{\Pi}_{\mu\nu}(k_0;x_1,y_1) \;=\;- \,e^2\,
	\int_{-\infty}^{+\infty} \frac{dp_0}{2\pi} \; {\rm tr}
	\left[\gamma_\mu \; \widetilde{\mathcal S}_F(p_0+k_0; x_1,y_1)
	\gamma_\nu \;
	\widetilde{\mathcal S}_F(p_0;y_1,x_1) \right] \;.
\end{equation}

By an analogous procedure to the one applied in the previous Section, the
form of $\widetilde{\mathcal S}_F$ may be obtained exactly, for
example, by expanding in powers of $g$, taking into account the form of the
singular term, and afterwards summing the resulting series. The outcome of
this procedure may be put in terms of the free-space fermion propagator,
${\mathcal S}_F^{(0)}(x_1,y_1)$ (we omit, for the sake of clarity, writing
the $p_0$ argument: it is the same in all the instances where it appears
here) as follows: 
\begin{equation}\label{eq:sf}
	\widetilde{\mathcal S}_F(x_1,y_1) \;=\;
\; \widetilde{\mathcal S}_F^{(0)}(x_1,y_1)  -\, 
\widetilde{\mathcal S}_F^{(0)}(x_1,0)  \, \frac{g}{ 1 + g
\, \widetilde{\mathcal S}_F^{(0)}(0,0)} \, \widetilde{\mathcal S}_F^{(0)}(0,y_1) \;.
\end{equation}

On the other hand, since ${\mathcal S}_F^{(0)}(x_1,y_1)$ 
may be shown to be given by:
\begin{equation}\label{eq:sf0}
	\widetilde{\mathcal S}_F^{(0)}(p_0;x_1,y_1) =\frac{1}{2} \; 
	[ - i \gamma_0 \sigma(p_0) \,+\,\gamma_1  \sigma(x_1-y_1) ]
	e^{-|p_0| |x_1-y_1|} \;,
\end{equation}
where $\sigma$ denotes the sign function, we can
render (\ref{eq:sf}) into a form which will be more convenient in our study
of $\Pi_{\mu\nu}$:
$$ \widetilde{\mathcal S}_F(p_0;x_1,y_1) =\frac{1}{2} \;\Big\{ 
- i \gamma_0 \sigma(p_0) \big[ e^{-|p_0| |x_1-y_1|} -
\frac{g^2 ( 1 - \sigma(x_1) \sigma(y_1))}{4+g^2} e^{-|p_0|(|x_1|+|y_1|)}\big]
$$
$$\,+\,\gamma_1 \big[ \sigma(x_1-y_1) \, e^{-|p_0| |x_1-y_1|} \, - \,
	\frac{g^2}{4 + g^2} (\sigma(x_1)-\sigma(y_1))
	e^{-|p_0|(|x_1|+|y_1|)}\big] \Big\}
$$
\begin{equation}
	  +\; \frac{g}{4 + g^2} \;\Big[ 
	 1 + \sigma(x_1) \sigma(y_1) + \gamma_5 \sigma(p_0) (
	\sigma(x_1) + \sigma(y_1)) \Big] \, e^{-|p_0|(|x_1| + |y_1|)}  \;.
\end{equation}

Introducing this expression for the propagator into (\ref{eq:pi}), one can
obtain $\widetilde{\Pi}_{\mu\nu}$. Note that because of the lack of
explicit Lorentz invariance (time and space coordinates are treated
differently), one should expect the calculation to miss a `seagull
term'~\cite{Treiman:1986ep}. Indeed, $\Pi_{\mu\nu}$ being the
correlator between current operators:
\begin{equation}
	\Pi_{\mu\nu}(x,y) \;=\; - \langle J_\mu(x) \, J_\nu(y) \rangle \;,
\end{equation}
a seagull term $\tau_{\mu\nu}(x,y)$ should be concentrated on $x=y$. Or, for the
partially Fourier transformed version, concentrated on $x_1=y_1$ and a
local polynomial in $k_0$. 

Coming back to the result for $\Pi_{\mu\nu}$, due to its dependence on the
sign of $x_1$ and $y_1$, we have found it convenient to present the result
according to the value of those signs: 

\begin{itemize}
	\item{$x_1 > 0$ and $y_1 > 0$}

In this case, we obtain for $\widetilde{\Pi}_{\mu\nu}(k_0;x_1,y_1)$ a
		result that may be written explicitly:
\begin{align}
	\widetilde{\Pi}_{\mu\nu}(k_0;x_1,y_1) \;=\; - \frac{e^2}{2\pi} \,
	|k_0| \,
	\Big\{
	\big[ e^{-|k_0| |x_1-y_1|} 
	+ \frac{g^2}{(1 + \frac{g^2}{4})^2}e^{-|k_0| |x_1+y_1|}\big] 
	\delta_{\mu 0} \delta_{\nu 0} \nonumber\\
      \,-\,\big[ e^{-|k_0| |x_1-y_1|} 
	- \frac{g^2}{(1 + \frac{g^2}{4})^2}e^{-|k_0| |x_1+y_1|}\big] 
	\delta_{\mu 1} \delta_{\nu 1} \nonumber\\ 
	\,+\, i \sigma(k_0) \, \big[ \sigma(x_1-y_1) \, e^{-|k_0| |x_1-y_1|} 
	- \frac{g^2}{(1 + \frac{g^2}{4})^2}e^{-|k_0| |x_1+y_1|}\big] 
	\delta_{\mu 0} \delta_{\nu 1}\nonumber\\
     	\,+\, i \sigma(k_0) \, \big[ \sigma(x_1-y_1) \, e^{-|k_0| |x_1-y_1|} 
	+ \frac{g^2}{(1 + \frac{g^2}{4})^2}e^{-|k_0| |x_1+y_1|}\big] 
	\delta_{\mu 1} \delta_{\nu 0} \Big\} \;.
\end{align}
\item{$x_1 > 0$ and $y_1 < 0$}

\begin{align}
\widetilde{\Pi}_{\mu\nu}(k_0;x_1,y_1) \;=\; - \frac{e^2}{2\pi} \,
	\frac{1 -  (\frac{g}{2})^2}{1 + (\frac{g}{2})^2} \; e^{-|k_0| |x_1-y_1|} \;
	\Big[ & |k_0| \big( \delta_{\mu 0} \delta_{\nu 0} \,-\, \delta_{\mu 1}
	\delta_{\nu 1} \big) \nonumber\\
	+ & i k_0 \big( \delta_{\mu 0} \delta_{\nu 1} + \delta_{\mu 1} \delta_{\nu 0} \big)
\Big] 
\;.
\end{align}
\end{itemize}
Note that, for the particular choice $g=2$, $\widetilde{\Pi}_{\mu\nu}$ vanishes,
exhibiting the decoupling from the current in the $x_1 > 0$ region from
the gauge field at $x_1 < 0$.
In other words, in this situation the induced current is insensitive to the
existence of a gauge field in the $x_1 < 0$ region.

The {\em form\/} of the vacuum polarization function in this case is identical, albeit suppressed by a $g$-dependent factor, to the one for a fermionic field in free spacetime. Besides, there is a covariantizing seagull term missing, due to
the lack of explicit Lorentz covariance in our calculation. Also, note that
that term should indeed be missing from the result obtained for $x_1>0$ and
$y_1<0$, which excludes $x_1=y_1$.

Denoting by $\widetilde{\Pi}^{(0)}_{\mu\nu}$ the VPT in the absence of the
wall, we recall the result for the vacuum polarization tensor corresponding
to the well-known result for the Schwinger model in the absence of borders,
with both arguments Fourier transformed,
\begin{equation}\label{eq:pi0}
	\widetilde{\Pi}^{(0)}_{\mu\nu} \;=\; \frac{e^2}{\pi} \; \big(
	\delta_{\mu\nu} - \frac{k_\mu k_\nu}{k^2} \big)  \;,
\end{equation}
we find that:
\begin{align}
\widetilde{\Pi}^{(0)}_{\mu\nu}(k_0;x_1,y_1) \;=\; - \frac{e^2}{2\pi} \,
 \; e^{-|k_0| |x_1-y_1|} \;
\Big[ & |k_0| \big( \delta_{\mu 0} \delta_{\nu 0} \,-\, \delta_{\mu 1}
\delta_{\nu 1} \big) \nonumber\\
	+  \, i k_0 \, \sigma(x_1-y_1) \,&\, \big( \delta_{\mu 0} \delta_{\nu 1} +
\delta_{\mu 1} \delta_{\nu 0} \big) \Big] \nonumber\\
	+ \, \frac{e^2}{\pi} \,\delta(x_1 - y_1)  \delta_{\mu 0}
	\delta_{\nu 0}\;,
\end{align}
where the last term is the seagull.
$\widetilde{\Pi}^{(0)}_{\mu\nu}(k_0;x_1,y_1)$ does of course satisfy the
Ward identity:
\begin{equation}
	i k_0 \widetilde{\Pi}^{(0)}_{0\nu}(k_0;x_1,y_1)\,+\, \partial_{x_1}
	\widetilde{\Pi}^{(0)}_{1\nu}(k_0;x_1,y_1) \,=\,0 \;.
\end{equation}
Indeed, it is just another version of 
$k_\mu \widetilde{\Pi}^{(0)}_{\mu\nu}=0$, which (\ref{eq:pi0}) clearly
satisfies.
When $g \neq 0$, it is straightforward to verify that also the part of the
VPT which depends on the presence of the wall, satisfies the Ward identity
by itself.
Therefore,
\begin{equation}
	i k_0 \widetilde{\Pi}_{0\nu}(k_0;x_1,y_1)\,+\, \partial_{x_1}
	\widetilde{\Pi}_{1\nu}(k_0;x_1,y_1) \,=\,0 \;.
\end{equation}

\subsection{Behaviour of the normal-current correlator} 
We study here the behaviour of $\widetilde{\Pi}_{\mu\nu}(k_0;x_1,y_1)$ when
$\mu = 1$ and one approaches the boundary with $x_1$: $x_1 \to 0$ (the
situation would be identical if one considered $\nu=1$ and $y_1 \to 0$,
instead).  This is the correlator between the normal component of the
current on the boundary, and both components of the current $J_\nu$. We
find, for $y_1 >0$:
\begin{equation}
	\widetilde{\Pi}_{1\nu}(k_0;0,y_1) \;=\; \frac{e^2}{2\pi} \,
	\Big[ 1 -  \frac{g^2}{(1 + \frac{g^2}{4})^2} \Big] \, e^{- |k_0| y_1} \, ( i
	k_0 \delta_{\nu 0} + |k_0| \delta_{\nu 1}) \;,
\end{equation}
and
\begin{equation}
	\widetilde{\Pi}_{1\nu}(k_0;0,y_1) \;=\; \frac{e^2}{2\pi} \,
\Big[ 1 -  \frac{g^2}{(1 + \frac{g^2}{4})^2} \big]  \, e^{|k_0| y_1} \, ( - i
	k_0 \delta_{\nu 0} + |k_0| \delta_{\nu 1}) \;,
\end{equation}
for $y_1 <0$. In both cases, the result vanishes only if $g=2$.

Finally, we note that $\widetilde{\Pi}_{\mu\nu}$ may be conveniently
written in an equivalent way:
\begin{equation}
	\widetilde{\Pi}_{\mu\nu}(k_0;x_1,y_1) \;=\; 
	\widetilde{\Pi}^{(0)}_{\mu\nu}(k_0;x_1,y_1) \,+\,
	\frac{g^2}{(1 + \frac{g^2}{4})^2} \, (-1)^\nu \,
	\widetilde{\Pi}^{(0)}_{\mu\nu}(k_0;x_1,-y_1) \;, 
\end{equation}
(no sum over $\nu$) in terms of $\widetilde{\Pi}^{(0)}_{\mu\nu}$. The second term proceeds from the
correlator between $J_\mu(x_0,x_1)$ and $J'_\nu(y_0,y_1) = (-1)^\nu
J_\nu(y_0,-y_1)$ (no sum over $\nu$), the current reflected on the wall.

\subsection{Vacuum polarization tensor in the presence of the
defect, in $3+1$ dimensions}\label{ssec:pi4}
Most of the previous results generalize in a rather straightforward way. Here, we use a Fourier representation where $x_\shortparallel \equiv
(x_0,x_1,x_2)$ are transformed, since there is translation invariance in
that hyperplane:
\begin{equation}
	{\mathcal S}_F(x;y) \;=\; \int \frac{d^3p_\shortparallel}{(2\pi)^3}
	e^{i p_\shortparallel \cdot (x_\shortparallel-y_\shortparallel)} \;
	\widetilde{\mathcal S}_F(p_\shortparallel;x_3,y_3) \;, 
\end{equation}
and we see that:
\begin{equation}\label{eq:pi4}
	\widetilde{\Pi}_{\mu\nu}(k_\shortparallel;x_3,y_3) \;=\;- \,e^2\,
\int \frac{d^3p_\shortparallel}{(2\pi)^3}
	\; {\rm tr} \left[\gamma_\mu \; \widetilde{\mathcal
	S}_F(p_\shortparallel +k_\shortparallel;x_3,y_3)
	\gamma_\nu \;
	\widetilde{\mathcal S}_F(p_\shortparallel ;y_3,x_3)
\right] \;.
\end{equation}

In this case, we will just present the most relevant result for
$\widetilde{\Pi}_{\mu\nu}(k_\shortparallel;x_3,y_3)$, namely, for the case
$x_3, y_3 > 0$.
Taking into account that ${\mathcal S}_F^{(0)}(x_3,y_3)$ 
is given by:
\begin{equation}
\widetilde{\mathcal S}_F^{(0)}(p_\shortparallel;x_3,y_3) =\frac{1}{2} \; 
	[ - i  \gamma_\shortparallel \cdot \hat{p}_\shortparallel \,+\,\gamma_3
	\sigma(x_3-y_3) ]
	e^{-|p_\shortparallel| |x_3-y_3|} \;,
\end{equation}
where $\hat{p}_\shortparallel \equiv
(\frac{p_\alpha}{|p_\shortparallel|})$, $\alpha = 0, 1, 2$, 
we find that (for $x_3, y_3 >0)$
\begin{equation}
\widetilde{\mathcal S}_F(p_\shortparallel;x_3,y_3) \;=\;
\widetilde{\mathcal S}_F^{(0)}(p_\shortparallel;x_3,y_3) 
\,+\, \frac{g/2}{1 + (\frac{g}{2})^2} \, 
\widetilde{\mathcal S}_F^{(0)}(p_\shortparallel;x_3,- y_3)  \gamma_3 
\;.
\end{equation}
Again, as it happened in $1+1$ dimensions, $\widetilde{\Pi}_{\mu\nu}$ may
be conveniently written is terms of the known, free-space result for the
VPT:
\begin{equation}
	\widetilde{\Pi}_{\mu\nu}(k_\shortparallel;x_3,y_3) \;=\; 
	\widetilde{\Pi}^{(0)}_{\mu\nu}(k_\shortparallel;x_3,y_3) \,+\,
	\frac{g^2}{(1 + \frac{g^2}{4})^2} \, e^{i \pi \delta_{\nu 3}} \,
	\widetilde{\Pi}^{(0)}_{\mu\nu}(k_\shortparallel;x_3,-y_3) \;,
\end{equation}
(no sum over $\nu$) which may also be thought of as the combination of a direct term and a suppressed image contribution.

\section{Conclusions and discussion}\label{sec:conc}
We have obtained the anomaly in the presence of a planar defect of a kind
that can impose bag boundary conditions in a special limit, showing that the anomaly is independent of the value of the coupling constant $g$.

Since the chiral anomaly is the same, away from the
defect, as the one for the free, no defect case ($g=0$). This has important
consequences in the $1+1$ dimensional case. To that end, we recall here
some well-known relations, which depend on that property, and apply them to
the case at hand: The axial current, $J_\mu^5\,\equiv\, i e \bar\psi
\gamma_\mu \gamma_5 \psi$ is, in $1+1$ dimensions, completely determined by
the vector current.  Indeed, in terms of expectation values,
\begin{equation}
	j_\mu^5 \;=\; \epsilon_{\mu \nu} j_\nu \;.
	\label{eq:acr1}
\end{equation}
Now, since the vector current is conserved and, because of the previous relation, 
its  curl is the anomaly, we have:
\begin{equation}\label{eq:divcurl}
\partial_\mu j_\mu \;=\; 0 \;\;,\;\;\;
\epsilon_{\mu\nu} \partial_\mu j_\nu \;=\; \frac{ie^2}{\pi} \epsilon_{\mu\nu}
	\partial_\mu A_\nu\;.
\end{equation}
Using the decomposition: 
\begin{equation}\label{eq:decomp}
j_\mu = \partial_\mu \varphi + i
\epsilon_{\mu\nu}\partial_\nu \chi \;,
\end{equation}
one finds:
\begin{equation}
	\partial^2 \, \chi \;=\; - \frac{e^2}{\pi} \epsilon_{\mu\nu}
	\partial_\mu A_\nu \;.  
\end{equation}
The general solution to the previous equation for $\chi$ may be written as
follows:
\begin{equation}\label{eq:chi}
	\chi \;=\; \chi_0 \,-\,\frac{e^2}{\pi} \,
	\frac{1}{\partial^2} \partial_\mu A_\nu 
\end{equation}
where, as usual, $\frac{1}{\partial^2}$ denotes the inverse of the
Laplacian in ${\mathbb R}^2$, with null conditions at infinity, and
$\chi_0$ is a harmonic function that enforces the boundary conditions. 
When one considers the $g=0$ case, $\chi_0 =0$, and inserting
(\ref{eq:chi}) into (\ref{eq:decomp}) one recovers the result for
$\Pi^{(f)}_{\mu\nu}$. In the $g = 2$ case, on the other hand, one considers
the $x_1>0$ region, and a non-trivial $\chi_0$ function is required in
order to make $j_1=0$ on the boundary. The result for $\Pi_{\mu\nu}$ in
this case may be interpreted as the one for $g=0$ plus a contribution
which can be understood as due to an image source, outside of the region.

Note that the combination of two effects: invariance of the anomaly and boundary conditions, completely determine the form of the VPT. 

\section*{Acknowledgements}
The authors thank CONICET, ANPCyT, and UNCuyo for financial support.

\end{document}